\pdfoutput=1

\documentclass[galaxies,article,accept,oneauthor,pdftex,12pt,a4paper]{mdpi}
\usepackage{graphicx}
\makeatletter
\renewcommand\@biblabel[1]{#1. }
\makeatother
\def\sun{\hbox{$\odot$}}
\def\h0units{\mathrm{km\,s^{-1}\,Mpc^{-1}}}

\def\nave{<n>\;}
\def\2F1{~_2F_1}
\newcommand{\om}{\Omega_{\rm M}}

\newcommand{\ok}{\Omega_K}
\newcommand{\ola}{\Omega_{\Lambda}}
\def\aap{A\&A\,  }
\def\aj{AJ  }
\def\apj{ApJ\,  }
\def\apjs{ApJS  }

\def\jcap{Journal of Cosmology and Astroparticle Physic  } 
 
\def\mnras{MNRAS\,  }

\def\pasp{PASP  }

\def\pra{Phys. Rev. A   }
\def\pre{Phys. Rev. E   }
\lastpage{x}
\doinum{10.3390/------}
\pubvolume{xx}
\pubyear{2015}
\history{Received: xx / Accepted: xx / Published: xx}
\Title{On the number of galaxies at high redshift}

\Author{Lorenzo Zaninetti $^{1,}$}

\address
{
$^{1}$
Dipartimento  di Fisica, via P.Giuria 1,\\ I-10125 Turin,Italy
}

\corres
{
zaninetti@ph.unito.it
}

\abstract
{
The number  of galaxies at a given flux  as  a function
of the redshift, $z$,   is derived when  the $z$-distance
relation is  non-standard.
In order to compare
different models, the same formalism is also applied
to the standard cosmology.
The observed luminosity function for galaxies
of the zCOSMOS catalog  at different redshifts
is modelled by
a new luminosity
function for galaxies, which is derived  by the truncated beta
probability density function.
Three  astronomical tests, which are the
photometric maximum as a function of the redshift for a fixed flux,
the mean value of the  redshift for a fixed flux,
and the luminosity function for galaxies as a function of the
redshift,
compare the theoretical values
of the standard and non-standard model
with the observed value.
The tests are  performed on the
FORS Deep Field (FDF) catalog  up to redshift $z=1.5$
and
on the zCOSMOS catalog extending  beyond  $z=4$.
These three tests show minimal differences
between the standard and the non-standard models.
}



\begin{document}


\section{Introduction}

The  linear correlation
between the expansion velocity, $v$, and $d_l$,
the distance in Mpc,    is
\begin {equation}
v= H_0 d_l  = c \, z
\quad  ,
\label{hubblevz}
\end{equation}
where
$H_0$    is the Hubble constant
$H_0 = 100 h \,\h0units $, with $h=1$
when  $h$ is not specified,
$c$ is  the  velocity of light,
$c=299792.458\mathrm{\frac{km}{s}}$
 see \citet{CODATA2012},
  and
$z$
is the redshift defined
as
\begin{equation}
z = \frac { \lambda_{obs} - \lambda_{em} } { \lambda_{em}} \quad ,
\end{equation}
with $\lambda_{obs}$  and
$\lambda_{em}$ denoting respectively
 the
wavelengths of the observed and emitted lines
as  determined from the lab source, the so called Doppler
effect.
This linear relation can be derived from
first principles, namely
general relativity (GR),
and is only a low-$z$ limit of a more general
relation in standard cosmology.
The presence of a velocity in the previous equation has pointed the standard cosmology towards an expanding universe.
The previous equation also contains a linear relation between
distance and redshift, and this has pointed the experts
in plasma physics towards an explanation of the redshift in terms
of the interaction of light with the electrons of the
intergalactic medium (IGM). In terms of plasma physics,
the expansion velocity of the universe becomes a pseudo-velocity
because the Universe is supposed to be Euclidean and static.
The conjecture here adopted is now outlined:
the redshift should be considered as a
spectroscopic measure independent
of the adopted cosmology.
An example of this conjecture
can be found at the home page
of the Supernova Cosmology Project (SCP)
`All the analyses
were developed with cosmology hidden.'
The main physical explanations for the redshift are:
a Doppler  shift, which means an expanding universe,
a general relativistic effect, see \cite{Kaiser2014},
a plasma effect, see
\cite{Brynjolfsson2004}, and
a tired light effect as  suggested by
\cite{Ashmore2006,Crawford2011}.
More details on the various  theories which
explain the  cosmological redshift can be found
in \cite{Marmet2009}.
{\it A point of discussion}: The presence of physical
effects which explain the redshift allows speaking of
  a Euclidean universe  in which the distances are evaluated
with the Pythagorean theorem.
In   a Euclidean  universe, the main parameters are $H_0$, $z$, and the distance, $d$,
 but
the velocity of expansion is virtual rather than real.
The aim   of having a   Euclidean  universe  is
the explanation of the astronomical variables such as
the redshift and the absolute magnitude and count
of galaxies without GR.


Concerning   the value of  $H_0$ and $\om$, we will adopt
recent values as obtained by
a  mixed model which uses the cosmic microwave background
(CMB) measurements at high redshift
and  the  baryon acoustic oscillation (BAO), see \cite{Bennett2014},
\begin{equation}
H_0 =(69.6 \pm 0.7 )\, \h0units
\quad ,
\end{equation}
\begin{equation}
\om  =0.286 \pm  0.008
\quad .
\end{equation}

Hubble's constant  is explained
in the dynamical relativistic  models beginning  with
\cite{Friedmann1922,Friedmann1924}.
Recently, research
in the framework of modern theories on an accelerating universe
has been focussed
on measuring cosmological
parameters such as $\om$  and $\ola$,
see  \cite{Riess1998,Perlmutter1999}.
In the last years, the enormous progress
in astronomical observations   has  increased
the available data   for galaxies up to $z=$3.36,
see the  FORS Deep Field  (FDF)  catalog, which is made
up of 300 galaxies with known spectroscopic redshift,
see \cite{Gabasch2004,Appenzeller2004}.
Another high redshift catalog  is  zCOSMOS,
which is made up of 9697 galaxies up to $z=$4, see \cite{Lilly2009}.
These   data demand a new formalism
for  the number of galaxies as a function of the redshift.
At the same time, Hubble's law can be inserted into a more precise
physical framework.
In order to cover these questions,
Section \ref{secbasic} reviews some old and new derivations
of Hubble's law as well the magnitude system,
and
Section \ref{secnonlinear} derives a new relation
for the number of galaxies as a function
of the  redshift.
Section \ref{relativistic} introduces a relativistic
model for the number of galaxies as a function of the redshift
and
Section \ref{evolution} deals with the luminosity
function for galaxies at different redshifts.

\section{Basic Formulae}

\label{secbasic}

\label{Hubble's law}

The change
of frequency of light in a gravitational framework
 \cite {Zwicky1929},
a photo-absorption process
between the photon  and the electron in  the intergalactic medium,
see Equation (3) in \cite{Ashmore2006},
or  a plasma effect,
see Equation (50) in
\cite {Brynjolfsson2004},
all give
\begin{equation}
v = c  \left[ \exp (H_0 \, d ) -1 \right]
\quad .
\label{ashmorevz}
\end{equation}

We can isolate the distance $d_l$ in Eq.~(\ref{hubblevz}),
obtaining  the linear relation
\begin{equation}
d_l=
{\frac {cz}{H_{{0}}}}
\quad .
\label{distancel}
\end{equation}
In standard cosmology, one cannot use the previous approximation
when inserting $d$ into
Eq.~(\ref{absolute}) for redshifts larger
       than  $\approx$ 0.1.
The previous equation
     needs to be generalised in standard cosmology
     when $z>0.1$
     with the use of both the  `Hubble function',
     see Eq.~(\ref{eq:ez}),
     and the  luminosity distance,
     see Eq.~(\ref{luminositydistance}).

\begin{figure*}
\begin{center}
\includegraphics[width=10cm]{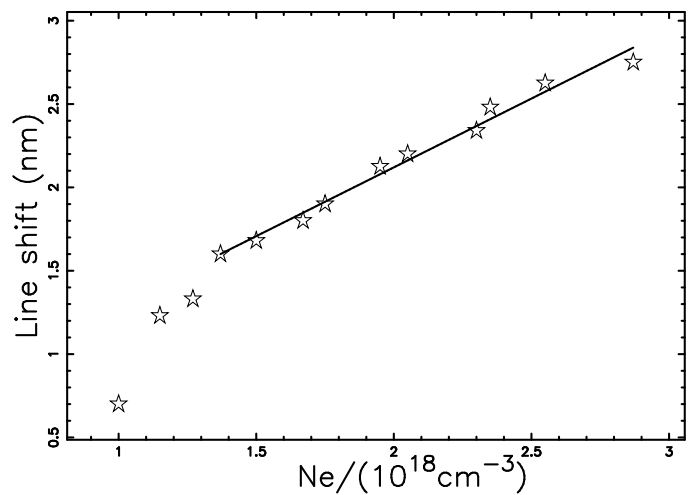}
\end {center}
\caption
{
HgI 435.83 nm line shifts
versus the electron density,
data as extracted  by the author
from   Figure 7 in \cite{Chen2009} (empty stars)
and linear regime                   (full line).
}
\label{highzlineshift}
    \end{figure*}

\begin{figure*}
\begin{center}
\includegraphics[width=10cm]{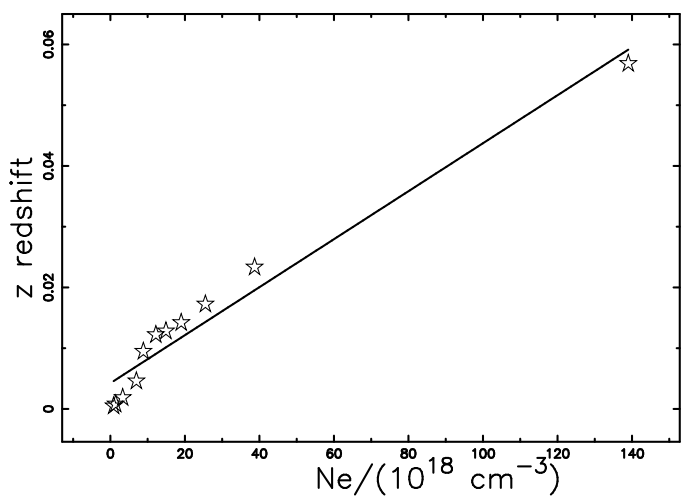}
\end {center}
\caption
{
Redshift of the $H\alpha$ line
versus the electron density,
data as extracted  by the author
from  Table 1   in \cite {Kielkopf2014} (empty stars)
and linear fit                          (full line).
}
\label{zhalfa}
    \end{figure*}

The expression  for the distance, $d$,
in the nonlinear Eq.~(\ref{ashmorevz})
gives the relation
\begin{equation}
d=
{\frac {\ln  \left( z+1 \right) c}{H_{{0}}}}
\quad .
\label{distancenl}
\end{equation}
A Taylor expansion around $z=0$  of this equation   gives
\begin{equation}
d=
{\frac {cz}{H_{{0}}}}
-\frac{1}{2}\,{\frac {c{z}^{2}}{H_{{0}}}}
+\frac{1}{3}\,{\frac {c{
z}^{3}}{H_{{0}}}}
\label{distancenltaylor}
\quad .
\end{equation}
In the limit  $z \rightarrow 0 $,  the linear distance, $d_l$,
and the nonlinear, $d$,
are equal.
The  laboratory results
of the line
shift in dense and hot plasmas
can be found in
\cite{Nguyen1986,Leng1995,Saemann1999,Zhidkov1999,WangYang2007}.
As  a first  example,
the experimental verification of the redshift of the
spectral line  of mercury as due to the surrounding
electrons      can be found in Figure  \ref{highzlineshift},
see also \cite{Ashmore2011}.
A second example is given by the
Balmer series $H\alpha$  line emitted
by laser-produced hydrogen.
A linear fit of the data in Table 1
in \cite{Kielkopf2014} gives
the  relation for the
redshift  of the $H\alpha$  line
\begin{equation}
z= 0.0042+0.00039 N_e
\quad when \quad 0.0042 < z <  0.0568
\quad ,
\end{equation}
where $N_e$ is the electron number density.
This linear relation can be
visualized in Figure \ref{zhalfa}.
The previous relation shows a laboratory redshift
as a function of a growing density.
In astrophysical plasma, we are interested in the redshift
as a function  of the distance when
the temperature of the medium  is 3K, as known
from the cosmic microwave background (CMB), and its mean
density is extremely low.
Therefore the previous laboratory experiment is illustrative rather
than quantitative.

\subsection{Magnitude system}

The  absolute magnitude of a galaxy, $M$,
is connected
with  the  apparent magnitude $m$ through the
relation
\begin{equation}
M = m - 5 \log_{10} (d) - 25
\quad .
\label{absolute}
\end{equation}
The  nonlinear absolute magnitude  as a function of the redshift
as given  by  the nonlinear  Eq.~(\ref{distancenl})  is
\begin{equation}
M =m  -5\, \log_{10}  \left( {\frac {\ln  \left( z+1 \right) c}{H_{{0}}}} \right)
-25
\quad .
\label{absolutenl}
\end{equation}
The previous formula predicts, from a theoretical
point of view, an upper limit on the maximum
absolute  magnitude which  can be observed in a
catalog of galaxies characterized by a given limiting
magnitude.
The previous curve can be connected with the
 Malmquist bias,
see \cite{Malmquist_1920,Malmquist_1922},
which was originally applied
to the stars and later on
to the galaxies by \cite{Behr1951}.
We now define the  Malmquist bias  as
the systematic distortion in luminosity or absolute magnitude  for the effective range of galaxies due to a failure in detecting those galaxies with  fainter luminosity or high absolute magnitude  at large distances.

\begin{figure}
\begin{center}
\includegraphics[width=7cm]{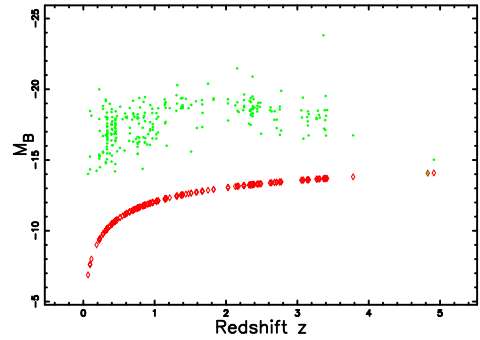}
\end{center}
\caption{
The blue absolute magnitude $M_B$
computed with the nonlinear Eq.~(\ref{absolutenl})
 for
263 galaxies belonging to the FDF  catalog
versus the well measured redshift.
The lower theoretical curve as represented by
the nonlinear Eq.~(\ref{absolutenl}) is  the
red thick line when $m_L$=30.33, which
is the maximum apparent magnitude
of the catalog,
$\mathcal{M_{\sun}}$ = 5.48  and
$H_{0}=69.6  \mathrm{\ km\ s}^{-1}\mathrm{\ Mpc}^{-1}$
(green points).
The redshift covers the range $[0,4.5]$
}
 \label{bias_fors}%
 \end{figure}

\begin{figure}
\begin{center}
\includegraphics[width=7cm]{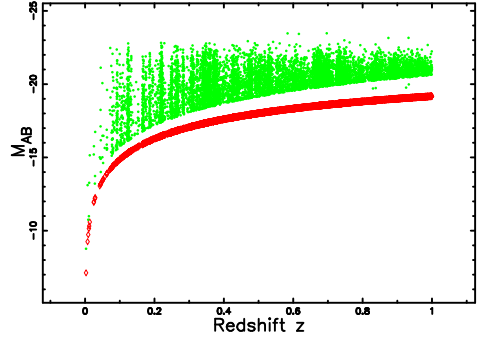}
\end{center}
\caption{
The B absolute magnitude $M$
computed with the nonlinear Eq.~(\ref{absolutenl})
for
9697 galaxies belonging to the zCOSMOS  catalog
versus the  redshift.
The lower theoretical curve as represented by
the nonlinear Eq.~(\ref{absolutenl}) is  the
red thick line when $m_L$=23.2,
$\mathcal{M_{\sun}}$ = 4.08 and
$H_{0}=69.6  \mathrm{\ km\ s}^{-1}\mathrm{\ Mpc}^{-1}$
(green points).
The redshift covers the range $[0,1]$
}
 \label{bias_zcosmos}%
\end{figure}

Figure \ref{bias_fors} shows such a curve, the Malmquist bias,
for the FDF catalog.
The  well measured spectroscopic  redshift,  the  blue  and
red   apparent magnitude
of the FDF data, can be found at
the Strasbourg Astronomical Data Centre (CDS).
Figure \ref{bias_zcosmos} shows the curve of the Malmquist bias
for the zCOSMOS catalog.
A careful examination of Figures \ref{bias_fors}
and \ref{bias_zcosmos} allows to conclude that all
the galaxies are in the region over the border line
of the Malmquist bias and the number of observed galaxies
decreases with increasing z  as theoretically predicted.

In a Euclidean, non-relativistic,
and homogeneous universe,
the flux of radiation without attenuation,
 $ f$,  expressed in units of $ \frac {L_{\sun}}{Mpc^2}$,
where $L_{\sun}$ represents the luminosity of the sun,
is
\begin{equation}
f  = \frac{L}{4 \pi d^2}
\quad ,
\end{equation}
where $d$   represents the nonlinear distance of the galaxy
expressed in Mpc, see Eq.~(\ref{distancenl}).
The relation connecting the absolute magnitude, $M$,
of a
galaxy  to  its luminosity is
\begin{equation}
\frac {L}{L_{\sun}} =
10^{0.4(M_{\sun} - M)}
\quad ,
\label{mlrelation}
\end{equation}
where $M_{\sun}$ is the reference magnitude
of the sun in  the considered bandpass.

The flux   expressed in units of $ \frac {L_{\sun}}{Mpc^2}$
as  a function of the  apparent magnitude is
\begin{eqnarray}
f=
7.957 \times 10^8 \,{e^{ 0.921\,{\it M_{\sun}}- 0.921\,{\it
m}}}
\quad    \frac {L_{\sun}}{Mpc^2} \quad ,
\label{damaf}
\end{eqnarray}
and  the inverse relation is
\begin{eqnarray}
m=
M_{\sun}- 1.0857\,\ln  \left(  0.1256 \times 10^{-8} f \right)
\quad .
\label{dafam}
\end{eqnarray}
Once a band  is fixed, we have a reference magnitude
of the sun in that band.
We give a few  examples:
band   $u^*$ in SDSS     $M_{\sun}$=6.38,
band     B   in FDF      $M_{\sun}$= 5.48 and
band     B   in zCOSMOS  $M_{\sun}$=4.08.

\subsection{Tired light}

\label{sectired}
Assume that the photon loses energy, $E$,
in a way proportional
to its energy:
\begin{equation}
\frac{dE}{dt} = -Cost\, E
\quad .
\end{equation}
The coefficient $Cost$  is assumed to be proportional
to the averaged number density of the IGM,
 $n$,
\begin{equation}
Cost = a n
\quad ,
\end{equation}
and therefore
\begin{equation}
\frac{dE}{dt} = -a n  E
\quad .
\end{equation}
We now replace the energy $E=h\nu$ where $h$ is Planck's
constant,
\begin{equation}
\frac{d\nu}{dt} = -a n  \nu
\quad ,
\end{equation}
and we convert the frequency to the wavelength,
\begin{equation}
\frac{d\lambda}{\lambda} = -a\, n\,  dt
\quad .
\end{equation}
The  speed of  light, $c$,
is assumed to be
constant, $\frac{ds}{dt}=c$
\begin{equation}
\ln{\frac{\lambda}{\lambda_0}}  = \frac{a}{c} \int_0^d n\, ds
\quad ,
\end{equation}
where $\lambda_0$ is the original wavelength.
On introducing the redshift
\begin{equation}
z=\frac{\lambda- \lambda_0}{\lambda_0}
\quad ,
\end{equation}
we obtain
\begin{equation}
\ln  (1+z) = \frac{a}{c} \nave \, d
\quad ,
\end{equation}
where $\nave$ is the averaged  number density.
The Hubble constant can be introduced with the following
meaning,
\begin{equation}
H_0 = a \nave
\quad ,
\end{equation}
obtaining
\begin{equation}
\ln (1+z) = \frac{H_0}{c} d
\quad .
\label{nonlzd}
\end{equation}
The distance  $d$
has   the same dependence on the photo-absorption/plasma  process
as  given  by Eq.~(\ref{distancenl})
\begin{equation}
d=
{\frac {\ln  \left( z+1 \right) c}{H_{{0}}}}
\quad .
\label{distancenltired}
\end{equation}
The observed flux without the absorption
of our galaxy  is
\begin{equation}
f  = \frac{L}{4 \pi d^2} \exp {(- \frac{H_0}{c} d)}
   =  \frac{L}{4 \pi d^2} \frac{1}{(1+z)}
\quad ,
\end{equation}
and  the  distance  modulus
for tired light is
\begin{equation}
m-M =
25+5\,{\frac {1}{\ln  \left( 10 \right) }\ln  \left( {\frac {\ln
 \left( 1+z \right) c}{{\it H_0}}} \right) }+\frac{5}{2}\,{\frac {\ln  \left( 1
+z \right) }{\ln  \left( 10 \right) }}
\quad  .
\label{modulustired}
\end{equation}
A generalization  of the concept of tired light  can be obtained by inserting
a supplementary absorption of  the light,
i.e., Compton scattering, see formula (51)
in \cite{Brynjolfsson2004}:
\begin{equation}
f  = \frac{L}{4 \pi d^2} \exp {(- \frac{H_0}{c} d
      -2 \frac{H_0}{c} d)}
   =  \frac{L}{4 \pi d^2} \frac{1}{(1+z)^{\beta}}
\label{fgeneralized}
\quad ,
\end{equation}
where $\beta$ is a variable parameter
which is  1 when  only tired light  is considered
and 3 when the Compton scattering is added.
Here we  have  invoked the Compton scattering
as a possible source of absorption
but  the parameter $\beta$ can be considered
a regulating parameter of an unknown scattering
mechanism.
The distance modulus  of generalized tired light
without  galactic extinction
is
\begin{equation}
m-M =
25+5\,{\frac {1}{\ln  \left( 10 \right) }\ln  \left( {\frac {\ln
 \left( 1+z \right) c}{{\it H_0}}} \right) }+\frac{5}{2}\,{\frac {\beta \ln  \left(
1+z \right) }{\ln  \left( 10 \right) }}
\label{modulusgeneralized}
\quad .
\end{equation}
Hubble's constant can be extracted from this equation
as a function of the distance modulus ($m-M$):
\begin{equation}
H_0 = 100000\,\ln  \left( z+1 \right) c{{\rm e}^{\frac{1}{2}\,
\beta\,\ln  \left( z+1
 \right) - \frac{1}{5}\, \left( m-M \right) \ln  \left( 10 \right) }}
 \quad \h0units
\quad,
\end{equation}
and, as a practical example, when $m-M$=43.834
and $z=0.974$, which is the case with SN C-001 in the  Union 2.1 compilation, we have $H_0 =69.6  \,\h0units $ when $\beta= 2.032$, which is the same value quoted in our Introduction.

A first comparison can  be done
with the distance modulus in a  plasma environment
as  given by  Eq.~(7) in \cite{Brynjolfsson2006}
without  galactic extinction:
\begin{equation}
m-M =
5\,{\frac {\ln  \left( \ln  \left( z+1 \right)  \right) }{\ln  \left(
10 \right) }}+\frac{15}{2}\,{\frac {\ln  \left( z+1 \right) }{\ln  \left( 10
 \right) }}+5\,{\frac {1}{\ln  \left( 10 \right) }\ln  \left( {\frac {
c}{H_{{0}}}} \right) }+25
\quad ,
\label{modulusplasma}
\end{equation}
see equation (7) in \cite{Brynjolfsson2006}.
A  second comparison can be done
with the historical  equation (23)
in  \cite{Sandage1961},
which  is the $[m,z]$ relation for the steady-model
\begin{equation}
m-M =
5\,{\frac {\ln  \left( z \right) }{\ln  \left( 10 \right) }}+5\,{
\frac {\ln  \left( 1+z \right) }{\ln  \left( 10 \right) }}+5\,{\frac {
1}{\ln  \left( 10 \right) }\ln  \left( {\frac {c}{{\it H_0}}} \right) }
+25
\quad  .
\label{modulussandage}
\end{equation}
We briefly recall that  Sandage \cite{Sandage1961} used
two explosive models, a Friedman model
\cite{Friedmann1922,Friedmann1924} and a steady state  model
\cite{Hoyle1956}.

\begin{figure*}
\begin{center}
\includegraphics[width=10cm]{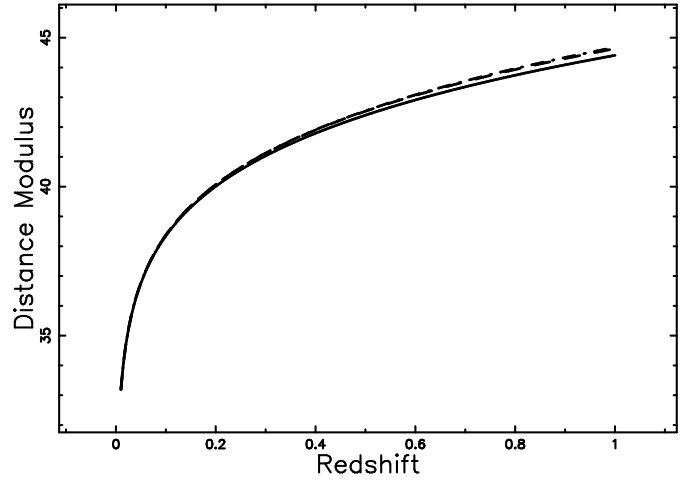}
\end {center}
\caption
{
Distance modulus  for the
generalized tired light
as given by Eq.~(\ref{modulusgeneralized}) (full line) with
$\beta$=2.7,
for  a plasma
as given by Eq.~(\ref{modulusplasma}) (dashed line),
and  for the steady-model
as given by Eq.~(\ref{modulussandage})
(dot-dash-dot-dash  line).
}
\label{modulus}
    \end{figure*}

Figure \ref{modulus}  presents the behavior
of the three distance moduli here considered
as functions of the redshift.
We now outline how
formula (\ref{distancenltired})
dates back to the year 1929.
The original formula
for the change
of frequency of light in a gravitational framework
is due  to  \cite{Zwicky1929}:
\begin{equation}
\frac {\Delta \nu} {\nu}  =
\frac{1.4  \pi G \rho D L}{c^2}
\quad .
\end{equation}
Here,
$\nu$  is the frequency,
$G$    is the Newtonian gravitational constant,
$\rho$ is the density in g/cm$^3$,
$D$    is the distance after which  the perturbing effect
       begins to fade out,
$L$    is the distance,  and
$c$    is the speed of light.

We make the change of variables
$\frac{1.4  \pi G \rho D L}{c} = a \nave$, $L=ds$,
${\Delta \nu}=-dE/h$, and $\nu=E/h$, which means
\begin{equation}
\frac{dE}{E} = - \frac{a}{c}  \nave \,ds
\quad .
\end{equation}
On repeating the previous  procedure, we obtain
formula (\ref{distancenltired}).

There is actually a debate on the existence  of tired light
in laboratory experiments. Here we use tired light
as a simple theory to model a more complex interaction
between light and the intergalactic plasma.

\subsection{The  luminosity function}

The  Schechter function,
introduced by
\citet{schechter},
provides a useful fit  for the  luminosity function (LF)  for galaxies:
\begin{equation}
\Phi (\frac{L}{L^*}) dL  = (\frac {\Phi^*}{L^*}) (\frac {L}{L^*})^{\alpha}
\exp \bigl ( {-  \frac {L}{L^*}} \bigr ) dL.
\label{equation_schechter}
\end{equation}
Here, $\alpha$ sets the slope for low values
of $L$, $L^*$ is the
characteristic luminosity, and $\Phi^*$ is a normalization.
The equivalent distribution in absolute magnitude is
\begin{equation}
\Phi (M)dM=0.921 \Phi^* 10^{0.4(\alpha +1 ) (M^*-M)}
\exp \bigl ({- 10^{0.4(M^*-M)}} \bigr)  dM \, ,
\label{lfstandard}
\end{equation}
where $M^*$ is the characteristic magnitude as derived from the
data.
The scaling with  $h$ is  $M^* - 5\log_{10}h$ and
$\Phi^* ~h^3~[Mpc^{-3}]$.

\section{$N$--$z$ relation}

\label{secnonlinear}

This section evaluates the number of galaxies
as a function of the redshift, firstly assuming
a linear relation
and secondly a nonlinear relation between redshift and distance.
The evaluations are done  on a sphere of radius $r$
which is identified with the chosen distance.
The main statistical test is
the $\chi^2$:
\begin{equation}
\chi^2 = \sum_{i=1}^n \left (\frac {T_i - O_i} {\sigma_i}\right)^2,
\label{chisquare}
\end{equation}
where $n  $   is the number of bins,
      $T_i$   is the theoretical value,
      $O_i$   is the experimental value represented
in terms of frequencies,
and  $\sigma_i$ is the  error computed as the square root
of  $O_i$.

\subsection{The linear case}

The joint distribution in {\it z}
and {\it f}  for galaxies,
see formula (5.133) in
 \cite{Peebles1993} or
 formula (1.104) in
 \cite{pad}
or
 formula (1.117)
in
\cite{Padmanabhan_III_2002},
 is
\begin{equation}
\frac{dN_l}{d\Omega dz df} =
4 \pi  \bigl ( \frac {c}{H_0} \bigr )^5    z^4 \Phi (\frac{z^2}{z_{crit}^2})
\label{nfunctionzschechter}
\quad ,
\end{equation}
where $d\Omega$, $dz$ and  $ df $ represent
the differential of
the solid angle,
the redshift, and the flux, respectively,
and     $\Phi$ is the Schechter LF.
The critical value of $z$,   $z_{crit}$, is
\begin{equation}
 z_{l,crit}^2 = \frac {H_0^2  L^* } {4 \pi f c^2}
\quad .
\end{equation}
The number of galaxies in $z$  and $f$ as given by
formula~(\ref{nfunctionzschechter})
has a maximum  at  $z=z_{l,pos-max}$,
where
\begin{equation}
 z_{l,pos-max} = z_{crit}  \sqrt {\alpha +2 }
\quad ,
\end{equation}
which  can be re-expressed   as
\begin{equation}
 z_{l,pos-max}(f) =
\frac
{
\sqrt {2+\alpha}\sqrt {{10}^{ 0.4\,{\it M_{\sun}}- 0.4\,{\it M^*}}}{
\it H_0}
}
{
2\,\sqrt {\pi }\sqrt {f}{\it c}
}
\quad  ,
\label{zmax_sch}
\end{equation}
or introducing the two observable variables, $M_{sun}$ and
 $M^*$,
\begin{equation}
 z_{l,pos-max}(m) =
\frac
{
1.772\times10^{-5}
\,\sqrt {2+\alpha}\sqrt {{10}^{ 0.4\,M_{{{\it \sun}}}-
 0.4\,{\it M^*}}}H_{{0}}
}
{
\sqrt {\pi }\sqrt {{{\rm e}^{ 0.921\,M_{{{\it \sun}}}-
 0.921\,m}}}{\it c}
}
\quad  .
\label{zmax_schmag}
\end{equation}
These two formulas, which model the photometric maximum, are
not reported in chapter 5 of \cite{Peebles1993}.

The theoretical mean redshift of galaxies with flux  $f$,
see formula~(1.105) in~\cite{pad},
 is
\begin{equation}
\langle z \rangle = z_{l,crit}  \frac {\Gamma (3 +\alpha)} {\Gamma (5/2 +\alpha)}
\quad .
\end{equation}

\subsection{The nonlinear case}

 We assume that  $f  = \frac{L}{4 \pi r^2}$
and
\begin{equation}
 z =e^{(H_0 \, r/c )} -1
\quad ,
\end{equation}
  where $r$ is the distance;
 in our case, $d$ is as represented
 by the nonlinear  Eq.~(\ref{distancenl}).
 The relation between $dr$
 and $dz$ is
\begin{equation}
dr = {\frac {c{\it dz}}{ \left( z+1 \right) H_{{0}}}}
\quad .
\end{equation}
The joint distribution in {\it z}
and      {\it f}  for the number of galaxies
 is
\begin{equation}
\frac{dN}{d\Omega dz df} =
\frac{1}{4\pi}\int_0^{\infty} 4 \pi r^2 dr \Phi (\frac{L}{L^*})
\delta\bigl(z-(e^{(H_0 \, r/c )} -1)\bigr)
\delta\bigl(f-\frac{L}{4 \pi r^2}    \bigr)
\quad ,
\label{nfunctionzschechternonldef}
\end{equation}
where $\delta$ is the Dirac delta function.

The evaluations of the integral  over luminosity and distances
gives
\begin{equation}
\frac{dN}{d\Omega dz df} = F(z;f,\Omega) =
\frac
{
4\, \left( \ln  \left( z+1 \right)  \right) ^{4}{c}^{5}{\it \Phi^*}\,
 \left( {\frac { \left( \ln  \left( z+1 \right)  \right) ^{2}}{{z_{{{
\it crit}}}}^{2}}} \right) ^{\alpha}{{\rm e}^{-{\frac { \left( \ln
 \left( z+1 \right)  \right) ^{2}}{{z_{{{\it crit}}}}^{2}}}}}\pi
}
{
{H_{{0}}}^{5}{\it L^*}\, \left( z+1 \right)
}
\quad  .
\label{nfunctionzschechternonl}
\end{equation}

The number of galaxies in $z$  and $f$ as given by
formula \ref{nfunctionzschechternonl})
has a maximum  at  $z=z_{pos-max}$,
where
\begin{equation}
 z_{pos-max}(f) =
{{\rm e}^{-1/16\,{\frac { \left( {\it Lstar}\,H_{{0}}-\sqrt {64\,\pi
\,\alpha\,{c}^{2}f+128\,\pi \,{c}^{2}f+{\it L^*}\,{H_{{0}}}^{2}}
\sqrt {{\it L^*}} \right) H_{{0}}}{\pi \,{c}^{2}f}}}}-1
\label{zmaximumflux}
\quad ,
\end{equation}
or introducing the two observable variables, $M_{sun}$ and
 $M^*$
\begin{eqnarray}
 z_{pos-max}(m) =&    \nonumber   \\
{{\rm e}^{-1/16\,{\frac { \left( {10}^{ 0.4\,{\it M_{sun}}- 0.4\,{\it
M^*}}H_{{0}}-\sqrt {64\,\pi \,\alpha\,{c}^{2}f+128\,\pi \,{c}^{2}f+
{10}^{ 0.4\,{\it M_{sun}}- 0.4\,{\it M^*}}{H_{{0}}}^{2}}\sqrt {{10}^
{ 0.4\,{\it M_{sun}}- 0.4\,{\it M^*}}} \right) H_{{0}}}{\pi \,{c}^{2
}f}}}}-1 & \quad .
\label{zmaximummagnitude}
\end{eqnarray}
The total number of galaxies
comprised between a minimum value of flux,
$f_{min}$, and a maximum value of flux $f_{max}$,
for the Schechter LF
can be computed through the integral
\begin{equation}
\frac{dN}{d\Omega dz} = \int_{f_{min}} ^{f_{max}}
F(z;f,\Omega)
df
\quad.
\label{integrale_schechter_tutte}
\end{equation}
This integral does not have an analytical expression
and we must perform
a numerical integration.

The theoretical mean redshift of galaxies with flux  $f$  can be deduced from
Eq.~(\ref{nfunctionzschechternonl}):
\begin{equation}
\langle z \rangle =\frac{\int_0^{\infty} z \,F(z;f,\Omega) dz}
{\int_0^{\infty} F(z;f,\Omega) dz}
\quad .
\label{zaverage}
\end{equation}
The above integral does not have an analytical expression,
and should  be numerically evaluated.
The differences  between the formulas of this
subsection and   the formulas in chapter 5  of
\cite{Peebles1993} are that our formalism
is built to cover high values of the redshift, against
the low values of the redshift of the standard approach.
Both models are built in the framework
of a Euclidean universe.

\section{Astrophysical applications}

We processed two catalogs in order to test the theoretical
formulae: the FDF and the zCOSMOS.
These two catalogs differ for the number of galaxies and parameters
of the Schechter luminosity function (LF)   for galaxies.
We now review the three parameters of the
Schechter  LF: $\alpha$ is fixed, $\Phi^*$ is not relevant because
we equalize the theoretical maximum in frequencies, and
$M^*$ is allowed to vary in order to match the observed frequencies.
The observed
mean redshift of galaxies with a flux  $f$ or $m$,
$\langle z_{obs} \rangle$,
is evaluated by the following algorithm.
\begin{enumerate}
\item A window in apparent magnitude or flux is
      chosen around $m$ or $f$.
\item All the galaxies which fall in the window
      are selected.
\item The mean value in redshift of $N$ selected galaxies is $\langle z_{obs} \rangle$ and the uncertainty in the
    mean, $\sigma_{\mu}$,  is $\sigma_{\mu} = s /\sqrt{N}$
    where $s$  is the standard deviation, see formula (4.14)
    in \cite{Bevington2003}.
\end{enumerate}

\subsection{The FDF catalog}

\begin{figure}
\begin{center}
\includegraphics[width=6cm]{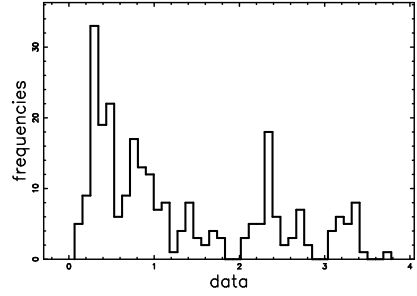}
\end {center}
\caption{
The galaxies  of the FDF
catalog are  organized in frequencies versus
spectroscopic   redshift.
The redshift  covers the range $[0,4]$ and the histogram's
interval is 0.1.
}
          \label{voids_fors}%
    \end{figure}

The pencil beam catalog FDF has a solid angle of $\approx$ 5.6  sq~arcmin,
or $7^{\prime} \times 7^{\prime}$  around the south galactic pole,
and covers the interval
$[0, 4]$ in redshift.
In particular, we selected the 263 galaxies with spectroscopical
redshift
and we processed the B band which has the range in apparent magnitude
 $[19.5, 30.2]$ mag.
 The reference magnitude for FDF in the  B band is
 $\mathcal{M_{\sun}}$ = 5.48.
The Schechter  LF for galaxies has  been
widely used to parametrise the LF in FDF
as a function of  the redshift.
As  an example in band B, $\alpha=-1.07$
in the range  $0.45 \leq z  \leq 1.11$
with  $-19.52 \leq  M^* \leq-18.80$,
see Figure 5  in \cite{Gabasch2004}.
We have maintained  $\alpha=-1.07$   but
we make $ M^*$ variable and specify it in the captions
of the figures.
The distribution of the spectroscopic
redshifts
in the FDF  is presented in Figure \ref{voids_fors}
and a comparison should be made
with the distribution of photometric redshifts,
see Figure 2 in \cite{Appenzeller2004}.

\begin{figure}
\begin{center}
\includegraphics[width=6cm]{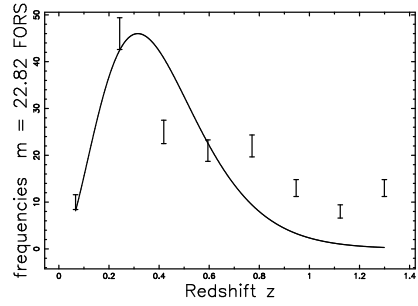}
\end {center}
\caption{
The galaxies  of the FDF  catalog   with
$ 22.08    \leq  m   \leq 26.81  $  or
$ 2.33   \frac {L_{\sun}}{Mpc^2} \leq
f \leq  181.39 \frac {L_{\sun}}{Mpc^2}$
are  organized in frequencies versus
spectroscopic   redshift.
The redshift covers the range $[0,1.5]$ and  the
histogram's interval  is 0.18.
The maximum frequency of observed galaxies is
at  $z=0.33$,
$\chi^2=$ 77.8, and the number of bins is 8.
The full line is the theoretical curve
generated by
$\frac{dN}{d\Omega dz df}(z)$
as given by the application of the Schechter LF
which  is Eq.~(\ref{nfunctionzschechternonl}) with
$\Phi^*=0.01 \,/Mpc^3 $, $M^*=-17.78$ and  $\alpha=-1.07$.
}
          \label{maximum_fors}%
    \end{figure}

\begin{figure}
\begin{center}
\includegraphics[width=6cm]{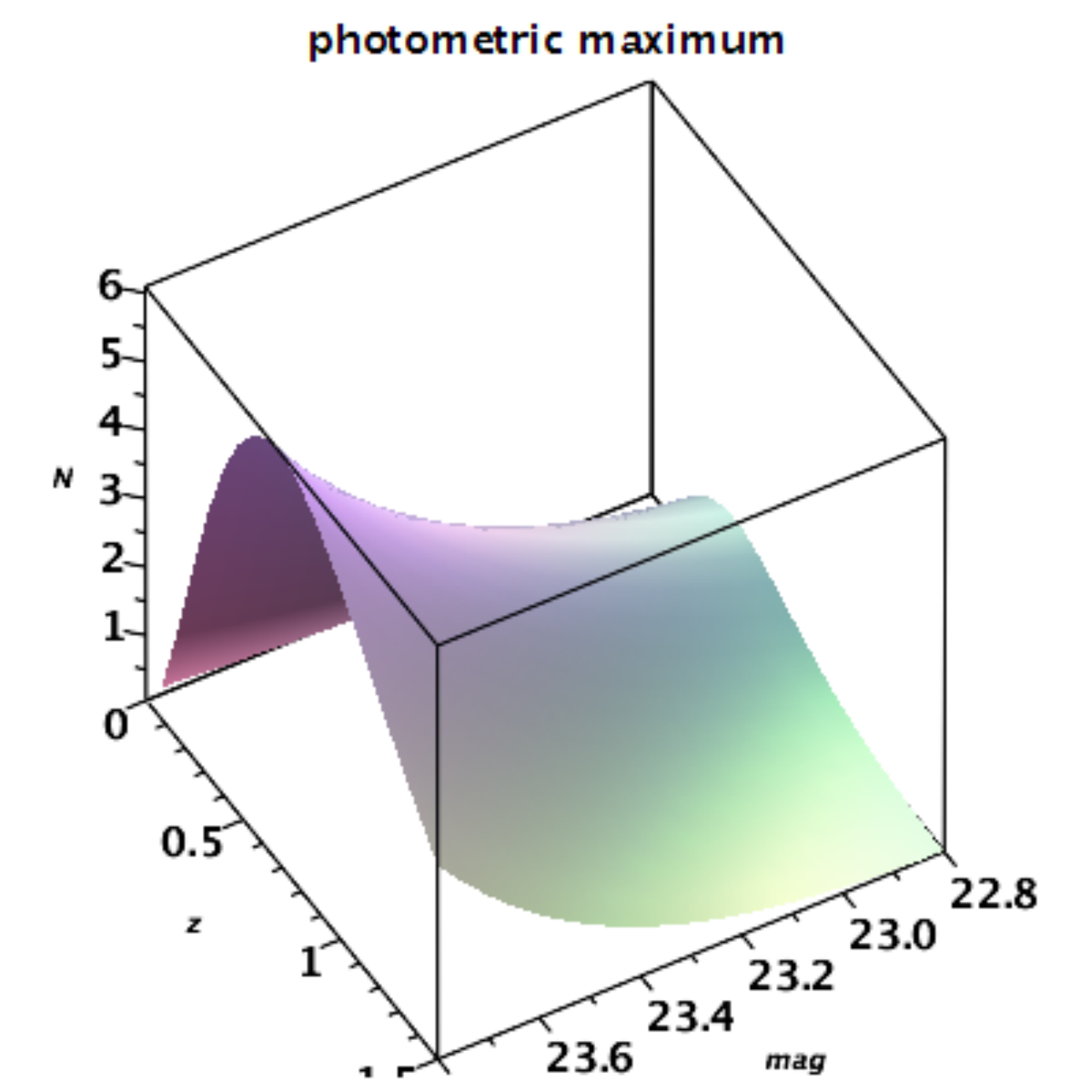}
\end {center}
\caption{
The  theoretical number of galaxies of the FDF catalog
as afunction of redshift and apparent magnitude
represented as a 3D surface, parameters
as in Figure \ref{maximum_fors}.
}
          \label{maximumzm}%
    \end{figure}

Figure \ref{maximum_fors}
presents the number of  observed  galaxies
in the FDF   catalog at a random
apparent magnitude and
Figure \ref{maximumzm} reports the theoretical number of galaxies 
as function of  redshift 
and apparent magnitude.   
Here we adopted the law of the rare events or Poisson distribution
in which the variance is equal to the mean, i.e.
the error bar is given by the square root of the frequency.
An enlarged discussion
on the validity of this approximation can be found
in \cite{Aggarwal2012}.

\begin{figure}
\begin{center}
\includegraphics[width=6cm]{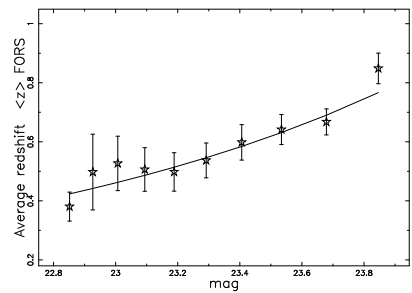}
\end {center}
\caption
{
Average observed redshift, $\langle z_{obs} \rangle$, as  a function
of the apparent magnitude for
the FDF catalog (empty stars)
and theoretical full line  for
$\langle z\rangle$
 as given
by the numerical integration of Eq.~(\ref{zaverage}).
Theoretical  parameters
as in Figure \ref{maximum_fors}.
}
          \label{zaverage_fors}%
    \end{figure}

A careful examination of Figure 3 in \cite{Appenzeller2004}
gives the maximum frequency of  galaxies with well
measured spectroscopic redshift in the FDF
at  $z \approx 0.3$.
The mean redshift of galaxies
as  a function of the apparent magnitude for the FDF catalog is presented
in   Figure \ref{zaverage_fors},
which shows an acceptable agreement between the data (empty stars)
and the theoretical values (full line).

\subsection{The zCOSMOS catalog}

\begin{figure}
\begin{center}
\includegraphics[width=6cm]{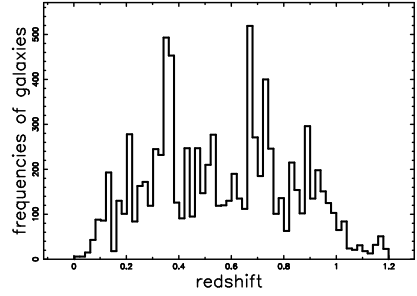}
\end {center}
\caption{
The galaxies  of the zCOSMOS  catalog
are  organized in frequencies versus
spectroscopic   redshift.
The redshift  covers the range $[0,1.2]$ and the histogram's
interval is 0.02.
}
          \label{voids_zcosmos_1}%
    \end{figure}

\begin{figure}
\begin{center}
\includegraphics[width=6cm]{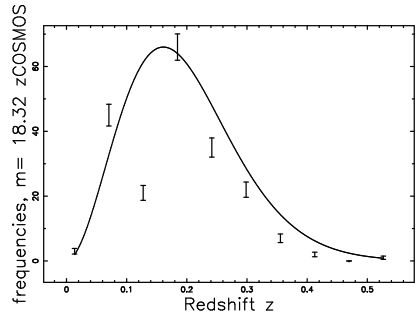}
\end {center}
\caption
{
The galaxies  of the zCOSMOS  catalog   with
$ 17.88   \leq  m   \leq 19.06 $  or
$ 803.43\frac {L_{\sun}}{Mpc^2} \leq
f \leq   2392.36 \frac {L_{\sun}}{Mpc^2}$
are  organized in frequencies versus
spectroscopic   redshift.
The redshift covers the range $[0,1]$ and the interval in the
histogram is 0.1.
The error bar is given by the square root of the frequency (Poisson
distribution) .
The maximum frequency of observed galaxies is
at  $z=0.213$,
$\chi^2=$ 147.3, and the number of bins is 10.
The full line is the theoretical curve
generated by
$\frac{dN}{d\Omega dz df}(z)$
as given by the application of the Schechter LF,
which  is Eq.~(\ref{nfunctionzschechternonl}) with
$\Phi^*=0.01 \,/Mpc^3 $, $M^*=-20.88$ and  $\alpha=-1.07$.
}
          \label{maximum_zcosmos}%
    \end{figure}

The zCOSMOS bright
redshift 10k catalog,
which covers a solid angle
$\Omega= 1.7 \deg^2$ or $\Omega=3.04617 \,10^{-4}$sr,
consists of 9697 galaxies in the
the interval
$[0, 4]$ in redshift
and range in the $I_{B}$ band
$[15, 23]$ mag, see \cite{Lilly2009}.
The reference magnitude for zCOSMOS in the  $I_{B}$  band is
 $\mathcal{M_{\sun}}$ = 4.08, see Table 2.1 in \cite{Binney1998}.
 The number of galaxies as a function of the redshift does not
have a continuous behavior: rather, we are in the presence of an
alternating behavior of voids and relative maxima,
see Figure \ref{voids_zcosmos_1}.
This nonhomogeneous spatial distribution of galaxies
can be made  continuous by introducing
bigger intervals in the computation of the frequencies,
e.g., a histogram
interval equal to 0.1.
Figure \ref{maximum_zcosmos}
presents the number of  observed  galaxies
in the zCOSMOS   catalog for  a given
apparent magnitude.

\begin{figure}
\begin{center}
\includegraphics[width=6cm]{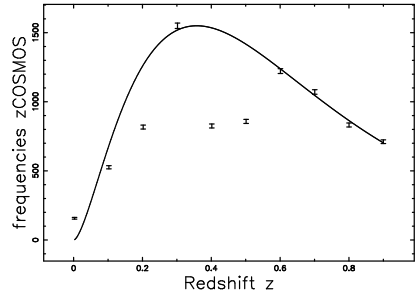}
\end {center}
\caption
{
All the  galaxies  of the zCOSMOS  catalog,
organized in frequencies versus
spectroscopic   redshift.
The redshift covers the range $[0,1]$ and the interval in the
histogram is 0.1.
The error bar is given by the square root of the frequency
(Poisson
distribution)
.
The maximum frequency of all observed galaxies is
at  $z=0.35$,
$\chi^2=$ 1864.65,  and the number of bins is 10.
The full line is the theoretical curve
generated by
$\frac{dN}{d\Omega dz}(z)$
as given by the numerical integration
of  Eq.~(\ref{integrale_schechter_tutte}) with
$\Phi^*=0.01 \,/Mpc^3 $, $M^*=-18$ and  $\alpha=-1.07$.
}
          \label{tutte_zcosmos}%
    \end{figure}

The total number of galaxies, $\frac{dN}{d\Omega dz}$,
can be computed
with the integral  represented  by
Eq.~(\ref{integrale_schechter_tutte}),
see Figure \ref{tutte_zcosmos}.

\begin{figure}
\begin{center}
\includegraphics[width=6cm]{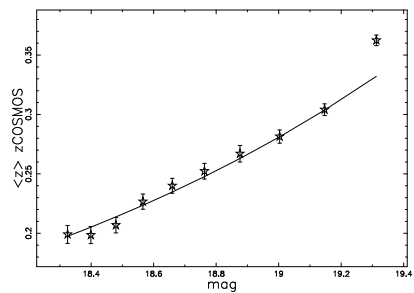}
\end {center}
\caption
{
Average observed redshift, $\langle z_{obs} \rangle$, as a function of
the apparent magnitude for the zCOSMOS catalog (empty stars)
and theoretical full line, $\langle z\rangle$,  as given
by the numerical integration of
Eq.~(\ref{zaverage}).
Theoretical  parameters
as in Figure \ref{maximum_zcosmos}.
}
          \label{zaverage_zcos}%
    \end{figure}

The mean redshift of galaxies
as  a function of the apparent magnitude for zCOSMOS is presented
in   Figure \ref{zaverage_zcos}.

\section{The relativistic case}

\label{relativistic}

The possibility of deriving an analytical result
for the number of galaxies as a function
of the redshift
in the relativistic case is connected with the availability
of an analytical expression for the luminosity distance.
We now use the same symbols as in  \cite{Hogg1999}
and we  define
the {\em Hubble
distance\/} $D_{\rm H}$
as
\begin{equation}
\label{eq:dh}
D_{\rm H}\equiv\frac{c}{H_0}
\quad .
\end{equation}
We then introduce a first parameter
 $\om$
\begin{equation}
\om = \frac{8\pi\,G\,\rho_0}{3\,H_0^2}
\quad ,
\end{equation}
where $G$ is the Newtonian gravitational constant and
$\rho_0$ is the mass density at the present time.
A second parameter is $\ola$
\begin{equation}
\ola\equiv\frac{\Lambda\,c^2}{3\,H_0^2}
\quad ,
\end{equation}
where $\Lambda$ is the cosmological constant,
see \cite{Peebles1993}.
The two previous parameters are connected with the
curvature $\ok$ by
\begin{equation}
\om+\ola+\ok= 1
\quad .
\end{equation}
The  comoving distance, $D_{\rm C}$,  is
\begin{equation}
D_{\rm C} = D_{\rm H}\,\int_0^z\frac{dz'}{E(z')}
\end{equation}
where $E(z)$ is the `Hubble function'
\begin{equation}
\label{eq:ez}
E(z) = \sqrt{\om\,(1+z)^3+\ok\,(1+z)^2+\ola}
\quad .
\end{equation}
The transverse comoving distance $D_{\rm
M}$ is
\begin{equation}
D_{\rm M} = \left\{
\begin{array}{ll}
D_{\rm H}\,\frac{1}{\sqrt{\ok}}\,\sinh\left[\sqrt{\ok}\,D_{\rm C}/D_{\rm H}\right] & {\rm for}~\ok>0 \\
D_{\rm C} & {\rm for}~\ok=0 \\
D_{\rm H}\,\frac{1}{\sqrt{|\ok|}}\,\sin\left[\sqrt{|\ok|}\,D_{\rm C}/D_{\rm H}\right] & {\rm for}~\ok<0
\end{array}
\right.
\end{equation}
An analytic expression for $D_{\rm M}$
can be obtained when
$\ola=0$:
\begin{equation}
D_{\rm M}=D_{\rm H}\,\frac{2\,[2-\om\,(1-z)-
(2-\om)\,\sqrt{1+\om\,z}]}{\om^2\,(1+z)}
~{\rm for}~\ola=0.
\end{equation}
A new form for $D_{\rm M}$ when $\ola=0$
is
\begin{equation}
D_{\rm M} = \frac
{
D_{{\rm H}}\sinh \left( 2\,{\rm arctanh} \left({\frac {1}
{\sqrt {1-\om }}}\right)-2\,{\rm arctanh} \left({\frac
{\sqrt {z\om +1}
}{\sqrt {1-\om }}}\right) \right)
}
{
\sqrt {1-\om}
}
\quad .
\end{equation}
The luminosity distance is
\begin{equation}
D_{\rm L} = (1+z)\,D_{\rm M}
\label{luminositydistance}
\end{equation}
which in the case
of
$\ola=0$
becomes
\begin{equation}
D_{\rm L} = (1+z)\,
\frac
{
D_{{\rm H}}\sinh \left( 2\,{\rm arctanh} \left({\frac {1}
{\sqrt {1-\om }
}}\right)-2\,{\rm arctanh} \left({\frac {\sqrt {z\om+1}
}{\sqrt {1-\om}}}\right) \right)
}
{
\sqrt {1-\om}
}
\quad ,
\label{relativisticlumdist}
\end{equation}
and  the distance modulus
\begin{equation}
m-M =
25+5\,{\frac {1}{\ln  \left( 10 \right) }\ln  \left( 2\,{\frac {c
 \left( 2-{\it \om}\, \left( 1-z \right) - \left( 2-{\it \om}
 \right) \sqrt {z{\it \om}+1} \right) }{H_{{0}}{{\it \om}}^{2}}}
 \right) }
\quad  .
\label{modulusrelativistic}
\end{equation}

We now return to $D_{\rm L}=r$:
the ration between the differential of the luminosity distance
and the differential of the redshift is
\begin{equation}
dr = \frac
{
c{\it \om}\, \left( 2\,\sqrt {{\it \om}\,z+1}+{\it \om}-2
 \right) {\it dz}
}
{
H_{{0}}\sqrt {{\it \om}\,z+1}{{\it \om}}^{2}
}
\quad .
\label{differentialrelativistic}
\end{equation}
This means  that  we have an analytical   expression
for
the differential
$dr=f(z)dz$  when
$\ola=0$.
This analytical differential will be inserted
later on in eqn.(\ref{nfunctionzschechternonldefrel}).
The  inverse relation between distance
and  redshift,
now denoted by
$z_M$,
is
\begin{equation}
z_M = \frac
{
rH_{{0}}{\om}^{2}+c{\om}^{2}-\sqrt {c{\om}^
{2} \left( 2\,rH_{{0}}+c \right) }\om-2\,c\om+2\,
\sqrt {c{\om}^{2} \left( 2\,rH_{{0}}+c \right) }
}
{
2\,c\om
}
\quad .
\end{equation}

\begin{figure}
\begin{center}
\includegraphics[width=6cm]{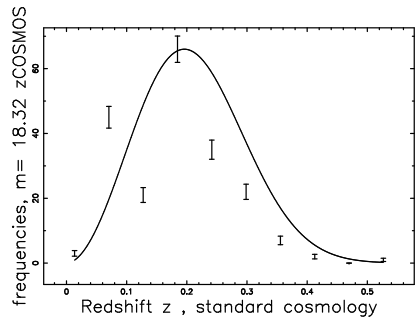}
\end {center}
\caption
{
The galaxies  of the zCOSMOS  catalog   with the same
parameters of Figure \ref{maximum_zcosmos}
are  organized in frequencies versus
spectroscopic   redshift.
The full line is the theoretical curve
generated by
$\frac{dN}{d\Omega dz df}(z)$
as given by the application of the Schechter LF
in the relativistic case,
which  is Eq.~(\ref{nzrelativistic}) with
$\Phi^*=0.01 \,/Mpc^3 $, $M^*=-20.7$, $\alpha=-1.07$
and  $\om=0.286$; $\chi^2$  =95.68  when the number of bin
is  10.
}
          \label{maximum_zcosmos_hogg}%
    \end{figure}

The joint distribution in {\it z}
and    {\it f}  for the number of galaxies
in the relativistic case   is
\begin{equation}
\frac{dN}{d\Omega dz df} =
\frac{1}{4\pi}\int_0^{\infty} 4 \pi r^2 dr \Phi (\frac{L}{L^*})
\delta\bigl(z-z_M\bigr)
\delta\bigl(f-\frac{L}{4 \pi r^2}    \bigr)
\quad ,
\label{nfunctionzschechternonldefrel}
\end{equation}
and its explicit value is
\begin{equation}
\frac{dN}{d\Omega dz df} = \frac{DNN}{DND}
\quad  ,
\label{nzrelativistic}
\end{equation}
where
\begin{eqnarray}
DNN =
64\,\Phi\, \left( \sqrt {\om \,+1}\om+\om \,-2\,
\sqrt {\om \,+1}-\om+2 \right) ^{4} \times
\\
\left( 4\,{
\frac { \left( \sqrt {\om \,+1}\om+\om \,-2\,
\sqrt {\om \,+1}-\om+2 \right) ^{2}}{{\om}^{4
}{z_{{{\it crit}}}}^{2}}} \right) ^{\alpha} \times
\\
{{\rm e}^{-4\,{\frac {
 \left( \sqrt {\om \,+1}\om+\om \,-2\,\sqrt {
\om \,+1}-\om+2 \right) ^{2}}{{\om}^{4}{z_{{{
\it crit}}}}^{2}}}}} \left( 2\,\sqrt {\om \,+1}+\om-2
 \right) {c}^{5}\pi
\quad,
\end{eqnarray}
and
\begin{equation}
DND =
{\om}^{9}\sqrt {\om z+1}{H_{{0}}}^{5}{\it L^*}
\quad .
\end{equation}
Figure \ref{maximum_zcosmos_hogg}
presents the number of  observed  galaxies
in the zCOSMOS   catalog for  a given
apparent magnitude in the relativistic case;
we adopted the value of $\om=0.286$
because
it is the concordance value, see \cite{Bennett2014}.

\begin{figure}
\begin{center}
\includegraphics[width=6cm]{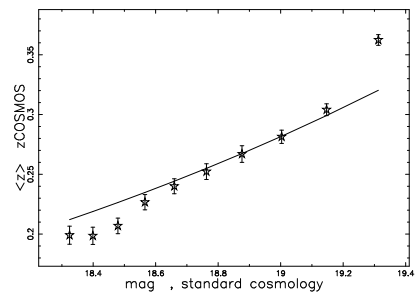}
\end {center}
\caption
{
Average observed  redshift, $\langle z_{obs} \rangle$, as  function of the apparent magnitude for  the zCOSMOS catalog (empty stars)
and theoretical full line, $\langle z\rangle$,   as given
by a numerical integration.
Theoretical  parameters
as in Figure \ref{maximum_zcosmos_hogg}.
}
          \label{zaverage_zcos_hogg}%
    \end{figure}

The mean numerical redshift of galaxies
as  a function of the apparent magnitude for zCOSMOS is presented
in   Figure \ref{zaverage_zcos_hogg} for the relativistic case.

The  observed galaxy number over-density
on cosmological scales up to second order in perturbation theory
with  all relativistic effects that arise
by  observing on the past lightcone are discussed
in \cite{Bertacca2014a,Bertacca2014b}.

\section{Evolutionary effects}

\label{evolution}
\begin{figure}
\begin{center}
\includegraphics[width=6cm]{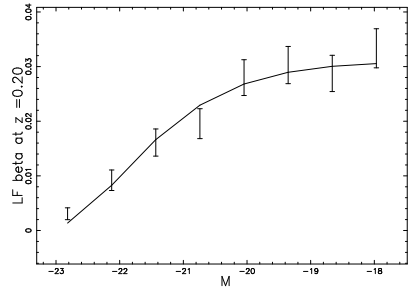}
\end {center}
\caption{
The luminosity function data of
zCOSMOS
are  represented with error bars.
The continuous line fit represents our beta LF
(\ref{lfmagnibeta}), the parameters are
$z=0.2$,
$\Delta z$=0.05
and  NDIV =8,
which means
$\chi^2  =5.35$.
}
          \label{evolution02}%
    \end{figure}

\begin{figure}
\begin{center}
\includegraphics[width=6cm]{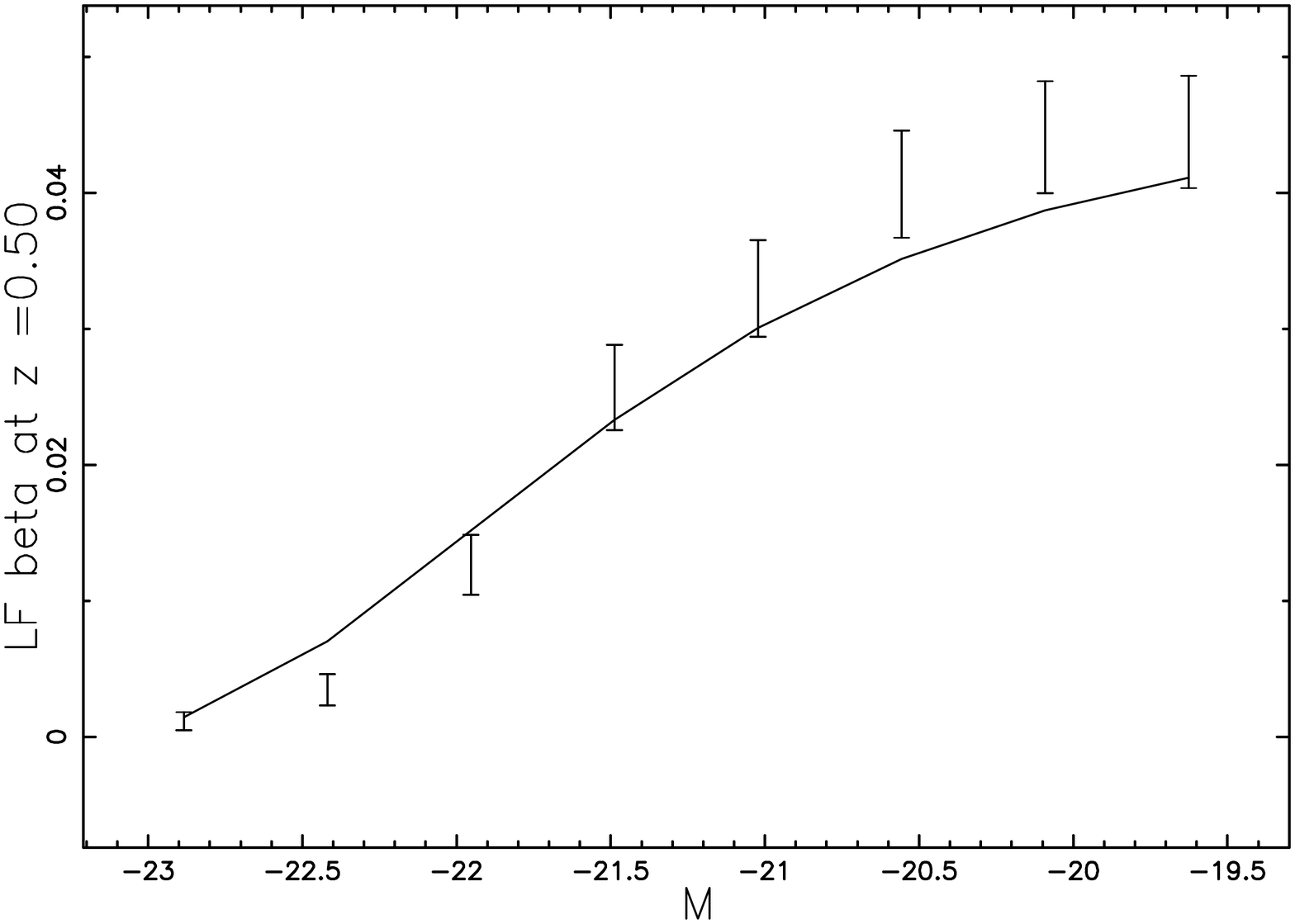}
\end {center}
\caption{
The luminosity function data of
zCOSMOS
are  represented with error bars.
The continuous line fit represents our beta LF
(\ref{lfmagnibeta}), the parameters are
$z=0.5$,
$\Delta z$=0.05,
and  NDIV =8,
which means
$\chi^2 $ =16.71.
}
          \label{evolution05}%
    \end{figure}

\begin{figure}
\begin{center}
\includegraphics[width=6cm]{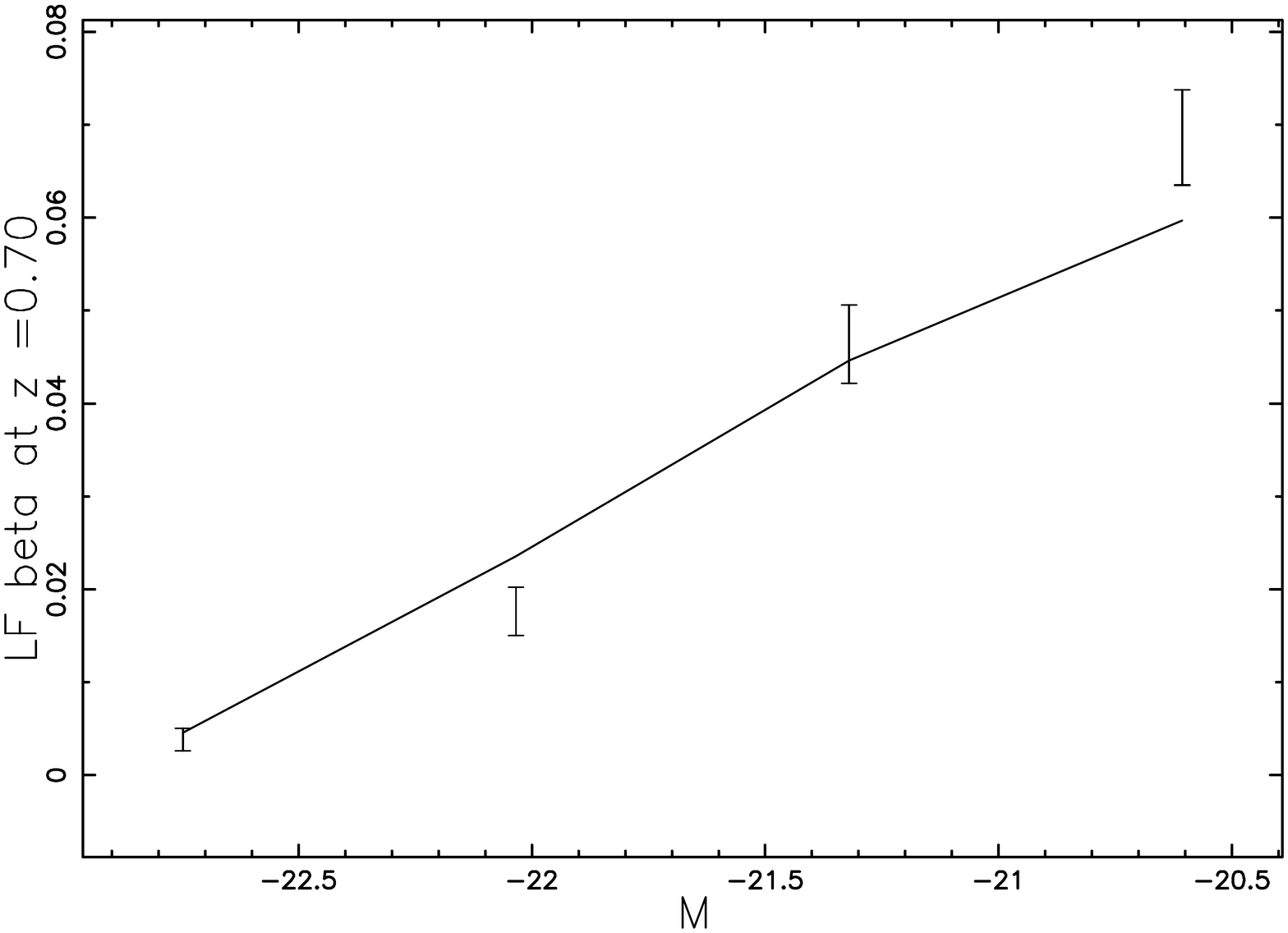}
\end {center}
\caption{
The luminosity function data of
zCOSMOS
are  represented with error bars.
The continuous line fit represents our beta LF
(\ref{lfmagnibeta}), the parameters are
$z=0.7$,
$\Delta z$=0.03,
and  NDIV =4,
which means
$\chi^2 $ =8.76.
}
          \label{evolution07}%
    \end{figure}

The main problem in modelling the
LF as a  function of the redshift is that the low
luminosity galaxies progessively disappear.
This observational fact can be solved
by adopting a truncated probability density function (PDF).
The beta distribution is
defined    in $[0, 1]$ and the  beta with scale  PDF
is defined in $[0, b]$.
On introducing a truncation in the beta PDF at low values,
we can model the observed LF as a function of the
redshift,    see
formula (34) in  \cite{Zaninetti2013a}.
Once the random variable $X$ is substituted
with the luminosity $L$,
we obtain  a new  LF  for galaxies,
$\Psi$,
\begin{equation}
\Psi (L) dL  =
K_L \, \Psi^*
{L}^{\alpha-1} \left( L_{{b}}-L \right) ^{\beta-1}dL
\quad , \label{betalf}
\end{equation}
where $\Psi^*$  is  a normalization factor which defines the
overall density of galaxies, a  number  per cubic Mpc.
The constant $K_L$ is
\begin{eqnarray}
K_L = \frac{A}{B}               \\
A   = &-\alpha\,\Gamma  \left( \alpha+\beta \right)\nonumber \\
B   = &{L_{{b}}}^{\beta-1}
{\mbox{$_2$F$_1$}(\alpha,-\beta+1;\,1+\alpha;\,{\frac {L_{{a}}}{L_{{b}}}})}
{L_{{a}}}^{\alpha}\Gamma  \left( \alpha+\beta \right)
\nonumber  \\
~&-{L_{{b}}}^{
\beta+\alpha-1}\Gamma  \left( 1+\alpha \right) \Gamma  \left( \beta
 \right)
           \quad ,      \nonumber
\end{eqnarray}
and  $L_a$, $L_b$ are   the lower, upper
values  in luminosity and
${\2F1(a,b;\,c;\,z)}$ is the
regularized hypergeometric
function
\cite{Abramowitz1965,NIST2010}.
The averaged luminosity,
$ { \langle L \rangle } $, is
\begin{eqnarray}
{ \langle L \rangle }
= K_L \, \Psi^*  \frac{A_N}{B_D} \quad ,
\end{eqnarray}
where
\begin{eqnarray}
A_N=&
-{L_{{b}}}^{\beta-1}{L_{{a}}}^{1+\alpha}
{\mbox{$_2$F$_1$}(-\beta+1,1+\alpha;\,2+\alpha;\,{\frac
{L_{{a}}}{L_{{b}}}})}
\Gamma  \left( 1+\alpha+\beta \right) \nonumber \\
~&
 +{L_{{b}}}^{\alpha+\beta}\Gamma
 \left( 2+\alpha \right) \Gamma  \left( \beta \right) \nonumber \\
B_D=&
\left( 1+\alpha \right) \Gamma  \left( 1+\alpha+\beta \right)
\quad . \nonumber
\end{eqnarray}
The relations connecting the
absolute magnitude $M$,
$M_a$ and  $M_b$
 of a
galaxy to the respective  luminosities  are
\begin{equation}
\frac {L}{L_{\sun}} =
10^{0.4(M_{\sun} - M)}
\; ,
\frac {L_a}{L_{\sun}} =
10^{0.4(M_{\sun} - M_a)}
\;
, \frac {L_b}{L_{\sun}} =
10^{0.4(M_{\sun} - M_b)}
\; ,
\label{magnitudes}
\end{equation}
where $M_{\sun}$ is the absolute magnitude
of the sun in the considered band.
The  beta truncated LF  in magnitude is
\begin{eqnarray}
\Psi (M) dM   =
-K_M \, \Psi^*
\,0.4\, \left( {10}^{- 0.4\,{\it am}+ 0.4\,{\it M_{\sun}}} \right) ^{
\alpha-1} \times & ~\nonumber \\
\left( {10}^{- 0.4\,{\it  M_b}+ 0.4\,{\it M_{\sun}}}-{10}^{-
 0.4\,{\it am}+ 0.4\,{\it M_{\sun}}} \right) ^{\beta-1}
 \times & ~ \nonumber \\
 {10}^{- 0.4\,{
\it am}+ 0.4\,{\it M_{\sun}}}\ln  \left( 10 \right) dM & ~
\quad ,
\label{lfmagnibeta}
\end{eqnarray}
where
\begin{eqnarray}
K_M   = \frac{A_M}{B_M}  &~               \\
A_M   = \alpha\,\Gamma    ( \alpha+\beta   )&~    \nonumber  \\
B_M   =
- \left( {10}^{- 0.4 {\it M_b}+ 0.4 {\it M_{\sun}}} \right) ^{\beta-1}
\times  &~
\nonumber \\
{\mbox{$_2$F$_1$}(\alpha,-\beta+1; 1+\alpha; {\frac {{10}^{- 0.4 {\it
M_a}+ 0.4 {\it M_{\sun}}}}{{10}^{- 0.4 {\it M_b}+ 0.4 {\it M_{\sun}}}}})}
\times  \nonumber &~\\
 \left( {10}^{- 0.4 {\it M_a}+ 0.4 {\it M_{\sun}}} \right) ^{\alpha}
\Gamma  \left( \alpha+\beta \right)
\nonumber &~ \\
 + \left( {10}^{- 0.4 {\it M_b}+
 0.4 {\it M_{\sun}}} \right) ^{\beta+\alpha-1}\Gamma  \left( 1+\alpha
 \right) \Gamma  \left( \beta \right)&~
\quad.
\nonumber
\end{eqnarray}
This  LF  contains the five parameters
$\alpha$,
$\beta$,
$M_a$,
$M_b$,
$\Psi^*$ which can be derived from the
operation of fitting
the observational data
and  $M_{\sun}$  which  characterize the considered band,
see \cite{Zaninetti2014d}.
The  number of variables  can be reduced to three
once $M_a$ and $M_b$ are identified with
the maximum and the minimum   absolute magnitude of  the
considered sample.
A further reduction of parameters can be realized
in the case of a well defined catalog of galaxies,
e.g., zCOSMOS, where
$M_a$=-23.47 mag, $\alpha=0.01$, $\beta=5$
and  $M_{\sun}$=4.08.
The low luminosity bound (high magnitude)
can be modelled in the  { \it classic case}
by extracting the
absolute magnitude from Eq.~(\ref{modulustired})
which represents the distance  modulus
for tired light by
\begin{equation}
M_b  =
{\it m_L}-\frac{5}{2}\,{\frac {\ln  \left( 1+z \right) }{\ln  \left( 10
 \right) }}-25-5\,{\frac {1}{\ln  \left( 10 \right) }\ln  \left( {
\frac {\ln  \left( 1+z \right) c}{H_{{0}}}} \right) }
\quad ,
\label{mbclassical}
\end{equation}
where $m_L$ is the limiting apparent magnitude, which
for zCOSMOS  is  $m_L=23.2$\ mag.
With the above choice of parameters, the observed LF  for zCOSMOS
as a function of the redshift  has only one free  parameter,
$\Psi^*$, which can be easily derived from
the fit of the histograms.
The observed LF for zCOSMOS can be built by adopting the
following algorithm.
\begin{enumerate}
\item A value for the redshift is fixed, $z$,  as well as the
      thickness of the layer, $\Delta z$.
\item All the galaxies comprised between $z$ and $\Delta z$
are selected.
\item The absolute magnitude can
be computed  from Eq.~(\ref{modulustired})
which represents the distance  modulus
for tired light.
\item The distribution in magnitude is organized in
  frequencies versus absolute magnitude.
\item The frequencies are divided by the volume,
    which is $V=\Omega  \pi r^2 \Delta r$,
    where $r$ is the considered radius, $\Delta r$ is the thickness
    of the radius, and $\Omega$ is the solid angle of ZCOSMOS.
\item The error in the observed LF is obtained  as
      the square root of the frequencies divided by the
      volume.
\end{enumerate}
Figures  \ref{evolution02}, \ref{evolution05}, and
\ref{evolution07} present  the LF of zCOOSMOS as well
the fit with the truncated   beta LF
at $z=0.2, z=0.5, and $z=0.7, respectively.

\begin{figure}
\begin{center}
\includegraphics[width=6cm]{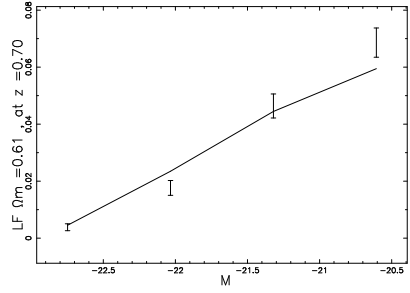}
\end {center}
\caption{
The luminosity function data of
zCOSMOS
are  represented with error bars.
The continuous line fit represents our beta LF
(\ref{lfmagnibeta}) in the relativistic case.
The input parameters are
$\ola=0$,
$\om=0.286$,
$z=0.7$,
$\Delta z$=0.03
and  NDIV =4, which means
$\chi^2 $ =8.78.
}
          \label{evolution07_hogg}%
    \end{figure}

In the relativistic case,
we can extract  the
absolute magnitude from
Eq.~(\ref {modulusrelativistic}),
which represents the distance  modulus
when $\ola=0$:
\begin{equation}
M_b  =
{\it m_l}-5\,{\frac {\ln  \left( 2 \right) }{\ln  \left( 10 \right) }}-
5\,{\frac {1}{\ln  \left( 10 \right) }\ln  \left( {\frac {c \left( 2-{
\it \om}\, \left( 1-z \right) - \left( 2-{\it \om} \right)
\sqrt {z{\it \om}+1} \right) }{H_{{0}}{{\it \om}}^{2}}} \right)
}-25
\quad .
\label{mbrelativistic}
\end{equation}
Figure  \ref{evolution07_hogg} presents  the LF of zCOOSMOS as well
as the fit with the truncated   beta LF
when  $z=0.7$ in the relativistic case.

\section{Conclusions}

{\bf Results:}

A nonlinear formulation of Hubble's law allows
determinaing  an old/new relation for
the distance as a function of the redshift,
see Equation (\ref{distancenl}).
This distance,  when  inserted in
the definition of the joint distribution in {\it z}
and {\it f}  for the  number of galaxies,
allows the determination of a new  $N-z$  relation,
see  Eq.~(\ref{nfunctionzschechternonl}).
Two photometric  tests were  done  for
263 galaxies belonging to the FDF  catalog
and for 9697 galaxies belonging to the zCOSMOS  catalog.
The first   test is dedicated  to the
photometric maximum in redshift, for which is possible to derive
an  analytical  expression  as a function of  the flux, see
Eq.~(\ref{zmaximumflux}),
or the apparent magnitude, see
Eq.~(\ref{zmaximumflux}); this results  in Figures
\ref{maximum_fors} and \ref{maximum_zcosmos}.

A second   test is dedicated  to the
average redshift, for which
a numerical integration of Eq.~(\ref{zaverage})
should be done;
it results in Figures  \ref{zaverage_fors}
and \ref{zaverage_zcos}.

The same formalism can also be applied to the
relativistic case
once the relativistic luminosity distance is  given,
see Eq.~(\ref{relativisticlumdist}).
In the standard cosmology or relativistic case the
the joint distribution in {\it z}
and {\it f}  for the  number of galaxies is given
by Eq.~(\ref{nzrelativistic}).
A comparison between the Euclidean and relativistic model
can be made on the photometric maximum as represented by
Figures \ref{maximum_zcosmos} and \ref{maximum_zcosmos_hogg}.
The  $\chi^2$ test gives
$\chi^2=147.3$ for the Euclidean
case,  as represented by Eq.~(\ref{nfunctionzschechternonldef})
and
$\chi^2=94.27$ for the relativistic
case
as represented by Eq.~(\ref{nzrelativistic}),
but large oscillations are present in the
observed frequencies and therefore the definitive answer
is remanded to future efforts.

The observed LF for galaxies can be modelled by a truncated
beta LF,  see Eq.~(\ref{lfmagnibeta}).
This new LF with an appropriate choice of parameters
has only one free parameter, which is
the number of galaxies per cubic Mpc,
$\Psi^*$,  and this  parameter decreases with the redshift.
The high magnitude  bound, $M_b$,
can be  modelled both by a Euclidean model
as given by Eq.~(\ref{mbclassical})
and by a relativistic model, $\ola=0$,
as given by Eq.~(\ref{mbrelativistic}).
As an example, the  third test
for the observed LF for galaxies
at $z=0.7$  gives
$\chi^2 $ =8.76  for the Euclidean case
and
$\chi^2 $ =8.78  for the relativistic case
when $\ola=0$ and  $\om=0.286$.

{\bf Generalizated tired light:}

The presence of the  factor $\beta$
for adjustable tired light,
see  Eqs.~(\ref{fgeneralized}) and
(\ref{modulusgeneralized}), poses the  problem of its
determination.
Eq.~(\ref{modulusgeneralized}), which represents
the distance modulus, can be calibrated on the database
of supernova (SN) of type Ia.
A careful  determination of $\beta$ and $H_0$ can
provide a better determination of the Malmquist
bias, as represented by  Figures
\ref{bias_fors} and
\ref{bias_zcosmos},  which  present a lack of galaxies
just above  the red lines.

{\bf  Tired light versus GR}

The new distance modulus as represented by
Eq.~(\ref{modulusgeneralized})
requires a  careful comparison
with the standard  LCDM.
In the case of LCDM with $\ola=0$,
an analytical solution for the
luminosity distance  exists
and allows the determination
of the relation between the differential of the distance
 and the differential of the redshift,
see Eq.~(\ref{differentialrelativistic}).

In the case of  $\ola \neq 0$, an analytical expression
for the luminosity distance does not exist,
and as a consequence
we do not have at the moment of writing
a relation between the differential of the distance
 and the differential of the redshift.
The use of the Pad\'e approximant
can produce analytical results for the luminosity distance,  see
\cite{Adachi2012,Wei2014};
this approximation will be the subject of future
research.

{\bf The cells}

The cellular structure of our universe,
see  as an example
Figure \ref{voids_zcosmos_1},
is now the greatest inconvenience
to the application of the continuous models
for the number of galaxies as a function of the redshift,
and perhaps an explanation for why the
theoretical lines do not fit the data,
as an example see Figure \ref{tutte_zcosmos}.
We briefly recall that there is an actual  debate
on the dimension of the universe
which is modelled by $N \propto R^D$,
where $N$ is the number of galaxies,
$R$ the radius of the considered sphere, and $D$ the dimension.
An homogeneous
universe  means $D=3$.
In the concordance model, D  makes a
transition to D=3 at scales between 40 and 100 Mpc,
see \cite{Bagla2008}.
An accurate analysis of 2MASS Photometric Redshift catalogue (2MPZ),
shows an
agreement with the standard cosmological model;  the
homogeneous regime is reached
faster than a class of fractal models with  D < 2.75,
see \cite{Alonso2015}.
As an example of non-homogeneity
the value $D=1.87$  has been reported in \cite{Zaninetti2012m}
where the 2MASS Redshift Survey  catalog was analysed.

\acknowledgments{Acknowledgments}

I am  grateful to the anonymous referee for  useful
suggestions which  have
changed the structure of the paper.


\end{document}